\documentclass[aip,reprint,nofootinbib]{revtex4-1}
\usepackage{amsmath}
\usepackage{graphicx}
\usepackage[section]{placeins}
\usepackage{amsmath}
\usepackage{amssymb}
\usepackage{amsfonts}
\usepackage{dsfont}
\newcommand{\ket}[1]{|#1\rangle}             
\newcommand{\op}[1]{\widehat{#1}}         
\begin{document}
\title{Laser-induced radial birefringence and spin-to-orbital optical angular momentum
conversion in silver-doped glasses}
\author{Jafar Mostafavi Amjad}
\author{Hamid Reza Khalesifard}
\affiliation{Department of Physics,
Institute for Advanced Studies in Basic Sciences (IASBS),
45137-66731 Zanjan, Iran}
\author{Sergei Slussarenko}
\author{Ebrahim Karimi}
\email{karimi@na.infn.it}
\affiliation{Dipartimento di Scienze
Fisiche, Universit\`{a} di Napoli ``Federico II'', Complesso
Universitario di Monte S. Angelo, 80126 Napoli, Italy}
\author{Lorenzo Marrucci}
\affiliation{Dipartimento di Scienze Fisiche, Universit\`{a} di
Napoli ``Federico II'', Complesso Universitario di Monte S. Angelo,
80126 Napoli, Italy}
\affiliation{CNR-SPIN, Complesso di Monte S.
Angelo, 80126 Napoli, Italy}
\author{Enrico Santamato}
\affiliation{Dipartimento di Scienze Fisiche, Universit\`{a} di
Napoli ``Federico II'', Complesso Universitario di Monte S. Angelo,
80126 Napoli, Italy}
\begin{abstract}
Samples of Ag$^+$/Na$^{+}$ ion-exchanged glass that have been
subject to intense laser irradiation may develop novel optical
properties, as a consequence of the formation of patterns of silver
nanoparticles and other structures. Here, we report the observation
of a laser-induced permanent transverse birefringence, with the
optical axis forming a radial pattern, as revealed by the
spin-to-orbital angular momentum conversion occurring in a probe
light beam. The birefringence pattern can be modeled well as
resulting from thermally-induced stresses arising in the
silver-doped glass during laser exposure, although the actual
mechanism leading to the permanent anisotropy is probably more
complex.
\end{abstract}
\maketitle
\noindent Metal-doped dielectrics may acquire interesting linear and
nonlinear optical properties, arising from the combination of the
dielectric transparency and of the metallic
surface-plasmon-polariton (SPP) response, particularly when the
metallic component aggregates in nanoparticles, clusters, or more
complex structures, giving rise to a kind of self-assembled
metamaterial.\cite{Kreibig:95,Abdolvand:05} The SPP resonances
depend critically on the metal cluster size, shape, and distribution
so that these materials can be effectively tailored to make novel
functional devices for optoelectronics and
telecommunications.\cite{Auxier:06,Gonella:96,Friberg:87}
Silver-ion-exchanged glass is one of the most promising materials,
due to its easy manufacturing and flexible properties. Different
techniques have been demonstrated to control the silver cluster
structure, size, distribution, and phase separation, most of which
are based on applying a strong DC electric field or on exposing the
materials to intense laser
illumination.\cite{Stietz:01,Miotello:01,Kaganovskii:01,Jin:03,Abdolvand:05,Nahal:06}.

Recently, Nahal et al.\ reported that initially isotropic
silver-doped glasses become birefringent under strong laser
illumination during preparation, with the optical axis lying along
the laser propagation direction.\cite{Nahal:07} In this paper, we
report the observation that laser exposure generates also a
significant transverse birefringence, with the induced optical axis
forming a radial pattern around the laser beam axis. This pattern
gives rise to optical depolarization effects occurring at small
deflection angle (in contrast to the large-angle conoscopic effects
reported in Ref.\ \onlinecite{Nahal:07}). We link these
birefringence effects to a partial conversion of the spin angular
momentum (SAM) of the incoming photons into orbital angular momentum
(OAM), a process known as spin-to-orbital angular momentum
conversion (STOC).\cite{Marrucci:06,Marrucci:11} As a consequence of
STOC, the SAM and OAM degrees of freedom of photons become entangled
and the polarization becomes spatially variant, giving rise to the
apparent depolarization of the light beam. It should be emphasized
however that this form of depolarization is not originated by random
dephasing, as in natural light sources, but it is a deterministic
phenomenon that preserves the overall light coherence.

The investigated samples were prepared by the well known
Ag$^+$/Na$^+$ ion exchange technique.\cite{Najafi:92,Nahal:07,Nahal:07a}
Soda-lime glass slides of dimensions 39$\times$25$\times$0.85 mm
merged into a 96:4 (weight ratio) molten mixture of NaNO$_3$ and
AgNO$_3$ at 400$^\circ$C for 4 hr. The chemical composition of the
glass was: SiO$_2$, 80\%; CaO, 9.41\%; Na$_2$O, 4.0\%; MgO, 3.3\%;
Al$_2$O$_3$, 2.2\%; K$_2$O, 0.41\%; S, 0.2\%; Fe$_2$O$_3$, 0.11\%;
P$_2$O$_5$, 0.11\%; others, 0.26 \% (percentages in weight). The
ion-exchanged glass samples were then irradiated with a multi-line
Ar$^+$ laser beam focused onto a spot with a 3 mm diameter. Three
samples were prepared with different exposure doses: 100 W/cm$^2$
for 1 s (S1), 30 W/cm$^2$ for 10 s (S2), and 100 W/cm$^2$ for 10 s
(S3). These silver-doped glasses are moderately absorbing in the
visible, so the laser beam leads to a strong local heating during
exposure and to the development of a radial thermal gradient towards
the center of the laser spot. This thermal gradient generates in
turn a mechanical stress leading, by elasto-optic effect, to a
cylindrically symmetric radial birefringence. In the case of a
Gaussian beam profile, the birefringence pattern induced by this
mechanism during exposure can be calculated
analytically.\cite{Khazanov:99,Mosca:10} For an initially isotropic
material such as glass, the local direction of the optical axis is
radially oriented along the temperature gradient and the local
optical birefringent retardation $\delta$ is radially symmetric and
given by
\begin{equation}\label{eq:delta}
    \delta(r)=\delta_0\left[1+\frac{1}{2}\left(\frac{r_0}{r}\right)^2\left({e^{-\frac{2r^2}{{r_0}^2}}-1}\right)\right],
\end{equation}
where $r$ is the radial coordinate, $r_0$ is the waist radius of the
laser beam and $\delta_0$ is the asymptotic phase retardation for
large radii. The latter will depend on the glass thermal, optical,
and elasto-optical properties and on the light power. In our case,
the actual mechanism leading to the radial birefringence must
clearly be more complicated than that described above, because
during laser irradiation the Ag$^+$ ions are known to form
nano-clusters which migrate around the beam axis, contributing to
the induced optical anisotropy. In addition, the laser-induced
thermal stresses can deform the surface of the glass sample,
producing a lensing effect. Once the laser light is turned off and
the sample cools down, the silver-particle distribution, stresses,
and surface deformation are frozen, and the radial birefringence
thus becomes persistent. A detailed model of these rather complex
effects is under investigation, but it is reasonable to assume that
Eq.~(\ref{eq:delta}) remains approximately valid as a
phenomenological model, with parameters $\delta_0$ and $r_0$ now
depending also on the laser intensity and exposure time and on the
Ag$^+$ ion concentration and mobility.

The main optical effect of the birefringence pattern given by
Eq.~(\ref{eq:delta}) can be simply obtained using the Jones operator,
$\op{U}$, acting on the light polarization. In the basis of the
circular polarizations, the Jones operator $\op{U}$ assumes the
simple form
\begin{eqnarray}\label{eq:action}
    \op{U}\cdot\ket{\pm}=\cos{\left[\frac{\delta(r)}{2}\right]}\ket{\pm}-i\,\sin{\left[\frac{\delta(r)}{2}\right]}\ket{\mp}\,e^{\pm2i\phi},
\end{eqnarray}
where $\ket{+}$ and $\ket{-}$ denote the left and right circular
polarizations, respectively and $\phi$ is the azimuthal angle in the
transverse plane. As we see, if the input light is circularly
polarized, the transmitted light will be the superposition of an
unmodified wave, with amplitude reduced by the factor
$\cos(\delta/2)$, and a new wave having opposite polarization
handedness, with amplitude factor $\sin(\delta/2)$. The latter also
exhibits a vortex phase factor $e^{\pm2i\phi}$, corresponding to an
OAM of $\pm2\hbar$ per photon, exactly balancing the variation of
SAM (i.e., from $\pm\hbar$ to $\mp\hbar$): this is just the STOC
process.\cite{Marrucci:06,Marrucci:11} In the following, for
brevity, the polarization-inverted component with nonzero OAM
(assuming that the input has zero OAM) will be referred to as the
``STOC component'' of the outgoing light.

The polarization of the STOC component is always orthogonal to the
polarization of the input beam, which explains the observed
depolarization of the transmitted light. Depolarization occurs for
any input polarization (linear or elliptical), since any
polarization can be decomposed into the left and right circular
components, each undergoing STOC to a different (opposite) OAM
value, so after STOC they do not add up coherently anymore. The STOC
efficiency is given by $\sin^2[\delta(r)/2]$. Therefore, complete
STOC is expected only for certain radii, determined by the Eq.\
$\delta(r)=(2n+1)\pi$ with $n$ integer, with partial STOC elsewhere.
Because in the present case $\delta(r)$ is not small generally, this
varying STOC efficiency should give rise to a radially-oscillating
intensity profile of the transmitted STOC beam.
\begin{figure}[h]
    \includegraphics[width=7cm]{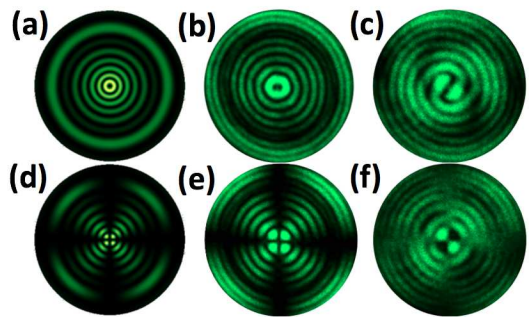}
    \caption{\label{fig:interference} (Color online) Far-field intensity profiles of a gaussian
    probe light beam passing through sample S3. Upper panels are for a left-circular
    input polarization, lower panels for linear polarization
    (vertical in the figures). Panels (a) and (d) are the profiles calculated
    from our theory, (b) and (e) are the experimental ones, and (c)
    and (f) are interference patterns with a gaussian reference
    beam. In the calculated patterns, we set $\delta_0=45$ and $r_0=0.7$ mm,
    and assumed a probe beam waist of 2 mm.}
\end{figure}

The optical properties acquired by our samples after the
laser-induced effects described above have become permanent were
studied using a probe laser beam having a gaussian input profile
with a beam waist diameter of about 4 mm, for two different input
polarizations: left-circular and linear. In both cases, in
transmission we observed a large far-field ring pattern, which is
typical of transverse phase-modulation, as observed also by other
authors and due to the isotropic laser-induced modulation of optical
properties.\cite{Kaganovskii:01} The depolarized light, however,
appears only in a small circular region around the pattern center,
that we selected by using an iris and a crossed polarizer.
Figures~\ref{fig:interference}b,e show the experimental intensity
patterns of this depolarized light, for two different input
polarizations. These patterns are very well reproduced by our
calculations of the STOC component based on our model, as shown in
Figs.~\ref{fig:interference}a,d. In the case of input circular
polarization, in particular, the intensity profile is cylindrically
symmetric and presents a dark singularity at the center,
characteristic of a beam carrying nonzero OAM. Moreover, the
patterns exhibit concentric circular fringes resulting from the
radial modulation of STOC efficiency, as discussed above. We also
recorded the interference patterns resulting from the superposition
of the depolarized light from the samples with a reference gaussian
beam (having the same polarization), in order to check the optical
phase distribution. The interference pattern obtained for a circular
polarization, in particular, presented the characteristic double
spiral pattern, confirming that the OAM eigenvalue is $|\ell|=2$, in
units of $\hbar$.\cite{Marrucci:06}

\begin{figure}[!htbp]
\begin{center}
    \includegraphics[width=5cm]{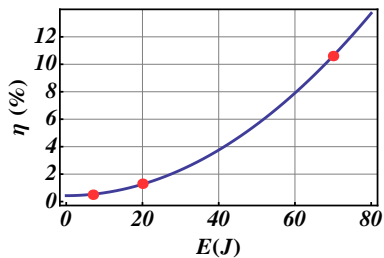}
    \caption{\label{fig:efficiency} (Color online) The STOC efficiency as a function
    of the total light energy deposited during the sample
    preparation. The circles correspond to the measured efficiency in our
    three samples S1, S2, and S3. The curve is a best fit with the quadratic expression
    $\eta(E)=a+bE^2$.}
\end{center}
\end{figure}

We investigated the behavior of the overall STOC efficiency (defined
as the power ratio between the space-integrated STOC component and
the overall transmitted light) for our different samples, having
different exposure parameters. Figure~\ref{fig:efficiency} shows
that the overall STOC efficiency is an increasing function of the
light energy dose absorbed by the sample during its preparation, as
could be reasonably expected.

The images shown in Fig.~\ref{fig:interference} provide a first
qualitative confirmation of the occurrence of STOC in our samples. A
more quantitative analysis can be performed by applying the methods
of quantum tomography to reconstruct the probe photon OAM quantum
state at the exit of the sample. This method allows us to retrieve
both the amplitudes and relative phases of all OAM components of the
transmitted light beam, despite the fact that they do not interfere.
Our model predicts that, for any input polarization, only
superpositions of OAM eigenstates with $\ell=\pm2$ should be
generated. Therefore, only the two-dimensional OAM Hilbert subspace
spanned by $\ell=\pm2$ is relevant. This space is isomorphous to the
polarization space, so we may use ``Stokes-like'' parameters to
describe it. The correspondence between the OAM and polarization
spaces can be made by associating the OAM eigenstates $\ell=\pm2$
with the left and right polarization states, respectively, and any
superposition of the OAM eigenstates $\ell=\pm2$ with the
corresponding elliptical polarization state. Thus, just as in the
case of polarization, to gain full information on the OAM state we
need only four independent measurements of the Stokes' parameters
$S_i$ $(i=0,\dots,3)$, from which the full density matrix of the
state can be retrieved.\cite{Padgett:99}
\begin{figure}[!htbp]
\begin{center}
    \includegraphics[width=6cm]{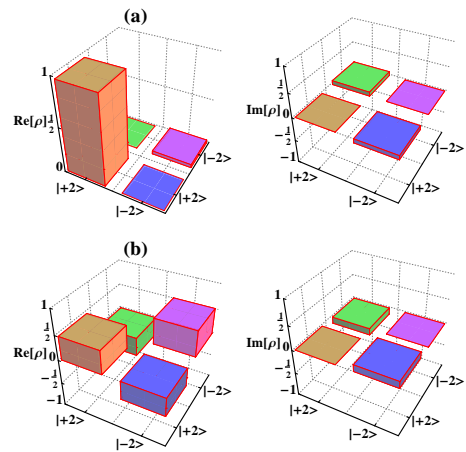}
    \caption{\label{fig:density_m} (Color online) Experimental density matrix for a
    (a) left-circularly polarized and (b) linearly polarized input probe beam,
    as emerging from sample S3. Left and right panels show the real and imaginary
    part of the density matrix, respectively (but, within the experimental errors,
    all measured density-matrix elements were found to be real).}
\end{center}
\end{figure}
To measure all OAM Stokes-like parameters we used six gray-scale
computer-generated holograms (CGH), as reported
elsewhere.~\cite{Nagali:09} These holograms were displayed in turn
on a spatial light modulator (SLM) on which the depolarized light
from the sample was made to impinge. The resulting far-field
intensity in the central area of the first-order diffraction was
then measured for each CGH and, by combining these data, all
Stokes-like parameters were calculated.~\cite{Nagali:09} The results
of this OAM photon state tomography for the two cases of
left-circular and linear input polarization of the probe beam are
shown in Fig.~\ref{fig:density_m}, which presents the real and
imaginary parts of the OAM density matrix. As we see, in the case of
the left circular input polarization, all output photons of the
depolarized component were put into the $\ell=+2$ OAM eigenstate
(confirming again the STOC effect), while in the case of the
vertical input polarization we obtained an equal-weight coherent
superposition of the two $\ell=\pm2$ OAM eigenstates. Moreover, the
fidelities of these states with the optical-field azimuthal profiles
of ideal $|\ell|=2$ and $\left(\ket{\ell=2}+\ket{\ell=-2}\right)/\sqrt{2}$ modes were found to be
$0.97$ and $0.91$, respectively. This confirms that the experimental
results are in excellent agreement with our STOC-based model.

In conclusion, we have reported the appearance of a radial
birefringence pattern in laser-exposed ion-exchanged silver-doped
glasses, as demonstrated by the occurrence of spin-to-orbital
angular momentum conversion in a probe beam. This pattern presumably
arises as a result of laser-induced thermal gradients and ensuing
silver nano-particle migration and permanently induced mechanical
stresses in the glass.

We acknowledge the financial support of the FET-Open program within
the 7$^{th}$ Framework Programme of the European Commission under
grant No.\ 255914, Phorbitech, and of IASBS research council under
the grant No.\ G2002IASBS101
%

\begin{thebibliography}{99}
%
\bibitem{Kreibig:95}  
U. Kreibig and M. Vollmer, {\it Optical properties of metal
clusters}, Springer series in materials science, Vol. 25 (Springer,
Berlin, 1995).

\bibitem{Abdolvand:05}  
A.~Abdolvand, A. Podlipensky, S. Matthias, F. Syrowatka, U. G\"osele, G. Seifert, and H. Graener,  Advanced Materials, {\bf 17}, 2983 (2005).

\bibitem{Auxier:06} 
J. Auxier, S. Honkanen, A. Sch\"ulzgen, M. Morrell, M. Leigh, S. Sen, N. Borrelli, and N. Peyghambarian, J.\ of the Opt.\ Soc.\ of Am.\ B, {\bf 23}, 1037
(2006).

\bibitem{Gonella:96}  
F. Gonella, G. Mattei, P. Mazzoldi, E. Cattaruzza, G. W. Arnold, G. Battaglin, P. Calvelli, R. Polloni, R. Bertoncello, and R. F. Haglund, Appl.\ Phys.\ Lett.\ {\bf 69}, 3101 (1996).

\bibitem{Friberg:87}  
S. Friberg and P. Smith, IEEE Journal of Quantum Electronics {\bf
23}, 2089 (1987).

\bibitem{Stietz:01}  
F.~Stietz, Appl.\ Phys.\  A-Materials Science \& Processing {\bf 72},
381 (2001).

\bibitem{Jin:03} 
R Jin, Y.  Cao, E. Hao, G. S. M\'etraux, G. C. Schatz, and C. A. Mirkin, Nature {\bf 425}, 487 (2003).

\bibitem{Nahal:06} 
A.~Nahal, J. Mostafavi-Amjad, A. Ghods, M. R. H. Khajehpour, S. N. S. Reihani, and M. R. Kolahchi, Journal of Appl.\ Phys.\, {\bf 100}, 1063 (2006).

\bibitem{Miotello:01} 
A.~Miotello, Appl.\ Phys.\ Lett.\ {\bf 79}, 2456 (2001).

\bibitem{Kaganovskii:01} 
Y.~Kaganovskii, I. Antonov, F. Bass, M. Rosenbluh, and A. Lipovskii, Journal of Appl.\ Phys., {\bf 89}, 8273
(2001).

\bibitem{Nahal:07} 
A.~Nahal and F.~Moslehirad, Journal of Materials Science, {\bf 42},
9075 (2007).

\bibitem{Marrucci:06} 
L.~Marrucci, C.~Manzo, and D.~Paparo, Phys.\ Rev.\ Lett. {\bf 96},
163905 (2006).

\bibitem{Marrucci:11} 
L.~Marrucci, E. Karimi, S. Slussarenko, B. Piccirillo, E. Santamato, E. Nagali, and F. Sciarrino, J. Opt.\ {\bf 13}, 064001 (2011).

\bibitem{Nahal:07a} 
A. Nahal, H. R. Khalesifard, Opt.\ Mat. {\bf 29}, 987, (2007).

\bibitem{Najafi:92}  
S.~I.~Najafi, {\it Introduction to Glass Integrated Optics}, edited
by S. I. Najafi (Artech House Publishers,1992).

\bibitem{Khazanov:99} 
E.~Khazanov, O.V. Kulagin, S. Yoshida, D.B. Tanner, and D.H. Reitze, IEEE Journal of Quantum Electronics {\bf 35},
1116 (1999).

\bibitem{Mosca:10} 
S. Mosca, B. Canuel, E. Karimi, B. Piccirillo, L. Marrucci, R. De Rosa, E. Genin, L. Milano, and E. Santamato, Phys.\ Rev.\ A {\bf 82}, 043806 (2010).

\bibitem{Padgett:99} 
M. Padgett and J. Courtial, Opt.\ Lett.,  {\bf 24}, 430 (1999).

\bibitem{Nagali:09} 
E. Nagali, F. Sciarrino, F. De Martini, B. Piccirillo, E. Karimi, L. Marrucci, and E. Santamato, Opt.\ Express, {\bf 17}, 18745 (2009).

\end{thebibliography}

%
\end{document}